\documentclass[twocolumn,aps,pra,showpacs,floatfix,superscriptaddress]{revtex4}

\usepackage{amsmath}
\usepackage{amsfonts}
\usepackage{amssymb}
\usepackage{bm}
\usepackage{bbm}
\usepackage{graphicx}
\usepackage{color}
\usepackage{maybemath}
\usepackage{fancybox}

\newcommand{\dd}{\mathrm{d}}
\newcommand{\ee}{\mathrm{e}}
\newcommand{\ii}{\mathrm{i}}

\newcommand{\calO}{\mathcal{O}}

\newcommand{\plus}{{\mbox{{\bf{\tiny +}}}}}

\newcommand{\wtilde}[1]{\mbox{$\widetilde #1$}}

\begin{document}

%
%
\title{Nonrelativistic Limit of the Dirac--Schwarzschild Hamiltonian:\\
Gravitational Zitterbewegung and Gravitational Spin--Orbit Coupling}

\author{U. D. Jentschura}
\affiliation{Department of Physics,
Missouri University of Science and Technology,
Rolla, Missouri 65409, USA}
\affiliation{MTA--DE Particle Physics Research Group,
P.O.Box 51, H--4001 Debrecen, Hungary}
\author{J. H. Noble}
\affiliation{Department of Physics,
Missouri University of Science and Technology,
Rolla, Missouri 65409, USA}

\begin{abstract} 
We investigate the nonrelativistic limit of the
gravitationally coupled Dirac Equation via a Foldy--Wouthuysen transformation.
The relativistic correction terms have immediate and obvious
physical interpretations in terms of a gravitational zitterbewegung, and a
gravitational spin-orbit coupling.  We find no direct coupling of the spin
vector to the gravitational force, which would otherwise violate parity.  
The particle-antiparticle symmetry described recently by one of us in
[Phys.Rev.A {\bf 87}, 032101 (2013)] is verified on the level of the perturbative
corrections accessed by the Foldy--Wouthuysen transformation.  The
gravitational corrections to the electromagnetic transition current are 
calculated.  
\end{abstract}


\pacs{11.10.-z, 03.70.+k, 03.65.Pm, 95.85.Ry, 04.25.dg, 95.36.+x, 98.80.-k}

\maketitle

%
%
\section{Introduction}
\label{sec1}

The ever-increasing precision of spectroscopic experiments necessitates the
consideration of gravitational effects in relativistic quantum mechanics.
General relativity effects already have to be taken into account in the
synchronization of clocks in the global positioning system (GPS) satellites,
which implies that the satellite-based clocks get ahead of ground-based clocks
by about 45\,$\mu$/d. For nonrelativistic neutron waves,
quantum-mechanical phase shifts due to the gravitational and inertial forces
have been measured in Refs.~\cite{CoOvWe1975,BoWr1983,BoWr1984}.  
The derivation of the gravitationally coupled Dirac equation 
has been discussed in textbooks~\cite{Iv1969a,Iv1969b,Mi1969,Go1985,%
IvMiVl1985,PeRi1987vol1,PeRi1987vol2}
as well as the original research literature~\cite{BrWh1957,Bo1975prd,%
SoMuGr1977,Re2011,Ye2011,Je2013,HeNi1990}.
Recently, it has been
argued that a symmetry exists~\cite{Je2013} 
for the gravitationally coupled Dirac equation which implies that
particles and antiparticles are both attracted to the gravitational field, and
that this symmetry holds to all orders in the velocity of the particles.
This symmetry was
obtained~\cite{Je2013} for a class of static space-time metrics which give rise
to a (generalized) Dirac--Schwarzschild
equation.  We refer to the corresponding Hamiltonian as the
Dirac--Schwarzschild Hamiltonian.

However, the result obtained in~\cite{Je2013} was not reconciled with
other articles from the literature, which investigate the nonrelativistic limit
of the quantum dynamics, and, in particular, discuss the conceivable
presence~\cite{Pe1978} of a spin-gravity coupling of the form $\vec \Sigma
\cdot \vec g$, where $\vec\Sigma$ is the $(4 \times 4)$-spin matrix, and $\vec
g$ is the acceleration due to gravity.  Specifically, in Ref.~\cite{Pe1978}, a
conceivable spin-gravity interaction and the pertinent experimental detection
have been investigated.  In Ref.~\cite{Ob2001}, via a Foldy--Wouthuysen
transformation~\cite{FoWu1950} of the Dirac--Schwarzschild Hamiltonian, a term
proportional to $\vec \Sigma \cdot \vec g$ is obtained in the leading
nonrelativistic approximation. It has been discussed in
Refs.~\cite{Ni2002,Ob2002} whether such a term would violate parity. Indeed, it
is well known that orbital as well as spin angular momenta constitute
pseudo-vectors.  Notably, the spin operator $\vec\Sigma = \gamma^5 \,\gamma^0
\, \vec\gamma$ transforms under parity as $\vec\Sigma \to \gamma^0 \,
\vec\Sigma \, \gamma^0 = \vec\Sigma$ and therefore as a pseudo-vector.  By
contrast, the gravitational force $\vec F_g = m \, \vec g$ with $m g = m \,
|\vec g| = G m M/r^2 = m \, r_s/(2 r)$ is a vector.  (Here, $m$ and $M$ are the
masses of the test particle and planet, respectively, and $r_s$ is the
Schwarzschild radius; we use natural units with $\hbar = c = \epsilon_0 = 1$).
One might thus conclude that any term in the Hamiltonian proportional to $\vec
\Sigma \cdot \vec g$ would indeed violate parity symmetry.  Apparently, the
question of how to physically interpret the leading relativistic corrections in
a curved space-time akin to the Schwarzschild geometry still constitutes an
open problem~\cite{SiTe2005,Si2008pra,ObSiTe2011,GoObSh2009}.

The gravitational Dirac Hamiltonian is similar in its mathematical structure to
the Dirac--Coulomb Hamiltonian~\cite{SwDr1991a,SwDr1991b,SwDr1991c}, and we
would a priori expect that the Foldy--Wouthuysen transformation should yield
similar terms, but respect the particle-antiparticle symmetry 
from~\cite{Je2013}. The
nature of the Foldy--Wouthuysen program is inherently perturbative.  In the
following, we expand to first order in the gravitational coupling constant,
i.e., we only keep terms of first order in the Schwarzschild radius $r_s$. The
corresponding dimensionless expansion parameter is $\chi = r_s/r$, where $r$ measures
the distance from the center of the black hole (in the sense of a space-time
coordinate). Regarding the momenta and distances, we assume that the Cartesian
components $p^i$ are of order $\xi m$ and that $r^i \sim 1/(\xi m)$, and expand to order
$\chi \, \xi^3$ or to order $\xi^4$ (for the gravitationally uncoupled terms).
After a rederivation of the Hamiltonian form of the 
Dirac--Schwarzschild equation in Sec.~\ref{sec2}, the Foldy--Wouthuysen
transformation is carried out explicitly in Sec.~\ref{sec3},
while conclusions are reserved for Sec.~\ref{conclu}.
We again reemphasize the use of natural units 
throughout the article ($\hbar = c = \epsilon_0 = 1$).

%
%
\section{Formalism}
\label{sec2}

We use the same conventions as in Refs.~\cite{Je2013} and assume that 
the curved-space Dirac $\gamma$ (overlined) and flat-space
(tilde) Dirac matrices fulfill the relations 
\begin{equation}
\label{gmunu}
\{ \overline \gamma^\mu(x), \overline \gamma^\nu(x) \} = 
2 \, \overline g^{\mu\nu}(x) \,,
\qquad
\{ \wtilde{\gamma}^\mu, \wtilde{\gamma}^\nu \} = 
2 \, \wtilde g^{\mu\nu} \,.
\end{equation}
Here, $\{ \cdot, \cdot \}$ is the anticommutator.
The curved-space metric is $\overline g^{\mu\nu}$ with $\mu,\nu =0,1,2,3$,
while the ``West-Coast'' flat-space metric is $\wtilde g^{\mu\nu} = {\rm
diag}(1,-1,-1,-1)$. Overlining the curved-space Dirac matrices
and using the tilde for the flat-space variants avoids 
a conceivable confusion with the  
particle physics literature~\cite{BjDr1964,BjDr1965,ItZu1980,PeSc1995,Sr2007},
where one denotes the flat-space matrices as $\gamma^\mu$,
and the flat-space metric as $g^{\mu\nu}$,
whereas in the literature on general relativity, 
one usually denotes the curved-space Dirac matrices 
without a tilde, and uses $g^{\mu\nu}$ for both 
flat-space as well as curved-space 
metrics~\cite{Di1975,Wa1984,Ch1992,Sc2009,Pa2010genrel}. 
Explicit use of the overlining and the 
tilde avoids any possible confusion.

The formulation of the gravitationally coupled
Dirac equation~\cite{Iv1969a,Iv1969b,Mi1969,Go1985,%
IvMiVl1985,PeRi1987vol1,PeRi1987vol2,%
BrWh1957,Bo1975prd,SoMuGr1977,Re2011,Ye2011,Je2013}
suggests the use of the vierbein formalism,
which is particularly straightforward to formulate
if the ``square root of the metric'' can be taken with ease
(see Sec.~6 of Ref.~\cite{Di1996}).
The Dirac action in curved space-time is given by the 
invariant integral
\begin{equation}
\label{symS}
S = \int \dd^4 x \; \sqrt{-\det \overline g} \;\; \overline\psi(x) \, 
\left( \frac{\ii}{2} \overline\gamma^\rho(x) \, 
\overleftrightarrow{\nabla}_\rho - m \right) \, \psi(x) \,,
\end{equation}
where the covariant derivative is $\nabla_\rho = \partial_\rho - \Gamma_\rho$,
and 
\begin{equation}
\label{solution}
\Gamma_\rho = - \frac{\ii}{4} \, \overline g_{\mu\alpha} \,
\left( \frac{\partial {b_\nu}^\beta}{\partial x^\rho} \,
{a^\alpha}_\beta - {\Gamma^\alpha}_{\nu \rho} \right) \,
\overline\sigma^{\mu\nu} \,.
\end{equation}
is the affine spin-connection matrix.
Here, the curved-space spin matrices are $\overline\sigma^{\mu\nu} =
\frac{\ii}{2} \, [ \overline \gamma^\mu, \overline \gamma^\nu ]$.
We represent the $\overline\gamma^\nu$ matrices in terms of the
flat-space $\wtilde\gamma^\mu$, 
\begin{subequations}
\label{ba}
\begin{align}
\overline\gamma_\rho =& \; {b_\rho}^\alpha \, \wtilde\gamma_\alpha \,,
\qquad
\wtilde\gamma_\rho = {a^\alpha}_\rho \, \overline\gamma_\alpha \,,
\\[0.007ex]
\overline \gamma^\alpha =& \; {a^\alpha}_\rho \, \wtilde\gamma^\rho\,,
\qquad
\wtilde \gamma^\alpha = {b_\rho}^\alpha \, \overline\gamma^\rho \,.
\end{align}
\end{subequations}
We use the flat-space Dirac matrices in the Dirac representation,
\begin{subequations}
\label{dirac_rep}
\begin{align}
\wtilde\gamma^0 =& \;
\left( \begin{array}{cc} \mathbbm{1}_{2\times2} & 0 \\
0 & -\mathbbm{1}_{2\times2} \end{array} \right) \,,
\quad
\wtilde\gamma^1 =
\left( \begin{array}{cc} 0 & \sigma^1 \\ -\sigma^1 & 0  \end{array} \right) \,,
\\[0.007ex]
\wtilde\gamma^2 = & \;
\left( \begin{array}{cc} 0 & \sigma^2 \\ -\sigma^2 & 0  \end{array} \right) \,,
\qquad
\wtilde\gamma^3 =
\left( \begin{array}{cc} 0 & \sigma^3 \\ -\sigma^3 & 0  \end{array} \right) \,,
\\[0.007ex]
\wtilde\gamma^5 =& \;
\ii \, \wtilde\gamma^0 \, \wtilde\gamma^1 \,
\wtilde\gamma^2 \, \wtilde\gamma^3
= \left( \begin{array}{cc} 0 & \mathbbm{1}_{2\times2} \\
\mathbbm{1}_{2\times2} & 0 \end{array} \right) \,.
\end{align}
\end{subequations}
The metric is recovered as
\begin{subequations}
\begin{align}
\{ \overline \gamma_\rho, \overline \gamma_\sigma\}
=& \; {b_\rho}^\alpha \, {b_\sigma}^\beta \,
\{ \wtilde\gamma_\alpha , \wtilde\gamma_\beta \}
= 2 \, \wtilde g_{\alpha \beta} \,
{b_\rho}^\alpha \, {b_\sigma}^\beta
= 2 \, \overline g_{\rho \sigma} \,,
\\[0.007ex]
\{ \overline \gamma^\rho, \overline \gamma^\sigma\}
=& \;
{a^\rho}_\alpha \, {a^\sigma}_\beta \,
\{ \wtilde\gamma^\alpha , \wtilde\gamma^\beta \}
= 2 \, \wtilde g^{\alpha \beta} \,
{a^\rho}_\alpha \, {a^\sigma}_\beta \,
= 2 \, \overline g^{\rho \sigma} \,.
\end{align}
\end{subequations}
The gravitationally coupled Dirac equation, obtained by variation 
of Eq.~\eqref{symS}, is
\begin{equation}
\label{gravdirac}
\left( \ii \, \overline\gamma^\mu \, \nabla_\mu - m \right) \psi(x) = 0 \,.
\end{equation}
We now specialize to a static space-time metric~\cite{Ed1924}
of a generalized Schwarzschild type, 
\begin{equation}
\label{vw}
{\overline g}_{\mu\nu} =
{\rm diag}\left( w^2(r), -v^2(r), -v^2(r), -v^2(r) \right) \,,
\end{equation}
where the vierbein coefficients are given as
${b_0}^\beta = {\delta_0}^\beta \, w(r)$,
${b_i}^j = {\delta_i}^j \, v(r)$,
${a^\alpha}_0 = {\delta^0}_\alpha/w(r)$, and
${a_i}^j = {\delta_i}^j/v(r)$.
The $a$ and $b$ matrices are symmetric in this case,
${a^\mu}_\nu = {a^\nu}_\mu$ and ${b_\mu}^\nu = {b_\nu}^\mu$.
With these coefficients, using computer algebra~\cite{Wo1999},
it is easy to evaluate all Christoffel symbols and to 
establish that~\cite{Je2013}
\begin{subequations}
\begin{align}
{\overline \gamma}^0 \, {\overline \gamma}^\mu \, 
\Gamma_\mu =& \; 
- \frac{{\wtilde \gamma}^0 \, \vec{\wtilde \gamma} \cdot \vec r}{r} \, G(r) \,,
\\[0.77ex]
G(r) =& \; \frac{ 2 \, v'(r) \, w(r) + v(r) \, w'(r)}{2 \, v^2(r) \,w^2(r)} \,.
\end{align}
\end{subequations}
The Hamiltonian form of the gravitationally
coupled Dirac equation,
\begin{equation}
\label{noncov}
\ii \left( {\overline \gamma}^0 \right)^2\,\partial_t \psi =
\left( {\overline \gamma}^0 \, {\overline \gamma}^j \, p^j +
\ii \, {\overline \gamma}^0 \, {\overline \gamma}^\mu \, \Gamma_\mu +
{\overline \gamma}^0 \, m \right) \, \psi \,,
\end{equation}
translates into $\ii \partial_t \psi = H \, \psi$,
where $H$ is given by 
\begin{equation}
H = \frac{w}{v} \, \vec \alpha \cdot \vec p
- \frac{\ii}{2 v} \, \vec \alpha \cdot \vec\nabla w
- \frac{\ii w}{v^2} \, \vec \alpha \cdot \vec\nabla v 
+ \beta m w \,.
\end{equation}
Here, we use the notation $\vec \alpha = 
{\wtilde \gamma}^0 \, \vec{\wtilde \gamma}$
and $\beta = {\wtilde \gamma}^0$. We now stretch space 
according to the scaling
\begin{equation}
\psi' = v^{3/2} \, \psi\,,
\qquad
H' = v^{3/2} \, H \, v^{-3/2} \,.
\end{equation}
This leads to a Hermitian Hamiltonian,
which acts on the Hilbert space of square-integrable
functions with the scalar product $\langle \phi, \psi \rangle = 
\int \dd^3 \, \phi^*(\vec r) \, \psi(\vec r)$.
in three-space reads as
\begin{equation}
H' = \frac12 \, \left\{ \vec\alpha \cdot \vec p, 
\mathcal F \right\} + \beta m w \,,
\qquad 
\mathcal F = \frac{w}{v} \,.
\end{equation}
We here confirm the result given in Eq.~(14) of 
Ref.~\cite{Ob2001}. For the Schwarzschild metric in 
isotropic coordinates (see Sec.~43 of Chap.~3 of Ref.~\cite{Ed1924}),
we have to first order in the Schwarzschild radius $r_s$,
\begin{align}
w =&\; 
\left( 1 - \frac{r_s}{4 r} \right) \, \left( 1 + \frac{r_s}{4 r} \right)^{-1} =
\frac{4 r - r_s}{4 r + r_s} \approx 1 - \frac{r_s}{2 r} \,,
\nonumber\\[1.0ex]
v =& \; \left( 1 + \frac{r_s}{4 r} \right)^2 \approx 1 + \frac{r_s}{2 r} \,,
\nonumber\\[1.0ex]
\frac{w}{v} =& \; \frac{16\,r^2\,(4 r - r_s)}{(4 r + r_s)^3} \approx 
1 - \frac{r_s}{r} \,.
\end{align}
The Schwarzschild radius is given as 
$r_s = 2 \, G \, M$, where $G$ is Newton's gravitational constant,
and $M$ is the mass of the planet.
So, to first order in $r_s$, we have to analyze the 
Dirac--Schwarzschild Hamiltonian $H_{\rm DS}$,
which is given by 
\begin{equation}
\label{HDS}
H_{\rm DS} =
\frac12 \, \left\{ \vec\alpha \cdot \vec p, 
\left( 1 - \frac{r_s}{r} \right) \right\} +
\beta m \left( 1 - \frac{r_s}{2 r} \right) \,.
\end{equation}
We can now carry through the Foldy--Wouthuysen program as usual.

%
%
\section{Foldy--Wouthuysen Transformation}
\label{sec3}

%
%
\subsection{Transformation of the Hamiltonian}

Contrary to widespread belief, the rationale of the Foldy--Wouthuysen
transformation~\cite{FoWu1950} actually is rather well-defined~\cite{BjDr1964}
and in some sense tied to the Dirac representation of the 
Dirac matrices given in Eq.~\eqref{dirac_rep}. One has to
(i)~identify the odd (in spinor space) part of the Hamiltonian $H$ and denotes
this odd part as~$\calO$.  (ii)~One then defines the Hermitian operator $S =
-\ii \, \beta \, {\calO}/(2 m)$ and the unitary operator $U = \exp(\ii S)$.
(iii)~The calculation of multiple nested commutators of $U$ and $H$ proceeds
until further nested commutators only yield higher-order terms when expressed
in terms of defined operational parameters of the expansion.  
(iv) If there are odd terms left after the first transformation, to the desired
order in the expansion, then one employs a second Foldy-Wouthuysen
transformation by identifying the odd part of the new Hamiltonian, which
resulted from the first Foldy-Wouthuysen step, as $\calO'$.
One does this recursively until all odd parts of
the input Hamiltonian are eliminated.  In the case of the free Dirac
Hamiltonian, it is possible to perform a Foldy--Wouthuysen transformation to
all orders in the momenta~\cite{BjDr1964},
but one may also choose a perturbative approach 
(see Appendix~\ref{appa}).  However, an exact
Foldy--Wouthuysen transformation has not been described for more complicated
Hamiltonians like the Dirac--Coulomb 
Hamiltonian~\cite{SwDr1991a,SwDr1991b,SwDr1991c}
\begin{equation}
\label{HDC}
H_{\rm DC} = \vec\alpha \cdot \vec p + 
\beta m - \frac{Z \alpha}{r} \,,
\end{equation}
where $Z$ is the nuclear charge number and $\alpha$ 
is the fine-structure constant,
or for other, nontrivial extensions of the standard 
free Dirac equation~\cite{YoAd2012}.
The Dirac--Schwarzschild Hamiltonian~\eqref{HDS} 
still is of an intricate nature.
A nonperturbative Foldy--Wouthuysen transformation 
has not been described in the literature for 
the Dirac--Coulomb Hamiltonian~\eqref{HDC}, and 
we can thus conclude that a perturbative approach 
seems most promising for the Dirac--Schwarzschild Hamiltonian.

For the gravitational correction terms, the expansions below are carried
through to to first order in $\xi= r_s/r$ and to 
third order in $\xi = \vec p/m$ or 
$\xi = 1/(m \, r)$, and we assume the exchanged photons to be soft,
i.e., $k \sim \xi^2 m$. This is the same expansion as for the 
Lamb shift~\cite{BjDr1964,Je1996} if we
identify $\chi$ with $Z\alpha$, where $Z$ is the nuclear
charge number, and $\alpha$ is the fine-structure constant. 
Terms are calculated up to order $\chi^4$ is
there is no gravitational interaction, and up to order $\xi \, \chi^3$ if there
is a gravitational interaction.  Terms of order $\xi^2$ (second order in the
gravitational interaction) are ignored.

To leading order in $r_s$, we have to analyze the Hamiltonian
$H_1 = H_{\rm DS}$ as given in Eq.~\eqref{HDS}.
For the first Foldy--Wouthuysen transformation, we therefore have
\begin{equation}
S_1 = -\frac{\ii}{2 m} \, \beta \, \calO_1 \,,
\qquad
\calO_1 = \frac12 \, \left\{ \vec\alpha \cdot \vec p, 
\left( 1 - \frac{r_s}{r} \right) \right\}  \,.
\end{equation}
The transformation is calculated as 
\begin{align}
\label{toH2}
H_2 =& \; \ee^{\ii S_1} \, H_1 \, \ee^{-\ii S_1} 
\nonumber\\[1.0ex]
=& \; H_1 + \ii \, [S_1,H_1] + 
\frac{\ii^2}{2!} \, [S_1, [S_1,H_1]] + \dots 
\end{align}
The result is
\begin{align}
H_2 =& \; \beta \left( m + \frac{\vec p^{\,2}}{2 m} 
- \frac{\vec p^{\,4}}{8 m^3} \right) 
- \beta \, \frac{m \, r_s}{2 \, r} 
\nonumber\\[1.0ex]
& \; + \beta \left( 
- \frac{3 r_s}{8 m} \, 
\left\{ \vec p^{\,2}, \frac{1}{r} \right\}
+ \frac{3 \pi r_s}{4 m} \delta^{(3)}(\vec r) \,
+ \frac{3 r_s}{8 m} \, \frac{\vec\Sigma \cdot \vec L}{r^3} \right) 
\nonumber\\[1.0ex]
& \; - \frac{(\vec \alpha \cdot \vec p)^3}{3 m^2} 
+ \frac14 \, \left\{ \frac{r_s}{r}, \vec\alpha \cdot \vec p \right\} \,.
\end{align}
A central ingredient of the Foldy--Wouthuysen transformation
is presence of the term $\beta\,m$ in the initial Hamiltonian
and the commutator relation $[ \beta \calO, \beta m] = -2 m\, \calO$,
which holds for any odd (in the spinor space) term $\calO$
in the Hamiltonian. Indeed, the first commutator in
equation ~\eqref{toH2} then eliminates the 
leading odd terms in the initial Hamiltonian $H_1$.
One might think that this scheme is not applicable
to the Hamiltonian~\eqref{HDS}, because the mass term is 
multiplied by a factor $[1-r_s/(2r)]$, 
but this is not the case: The nature of the Foldy--Wouthuysen 
transformation is perturbative, and the 
term $-\beta \, m \, r_s/(2r)$ is a perturbative gravitational
correction to the mass. The modification of the 
mass term therefore is of higher order in $\chi$ and does not inhibit the
application of the perturbative approach to the Foldy--Wouthuysen
transformation.

For the second Foldy--Wouthuysen transformation, we have
\begin{equation}
S_2 = -\frac{\ii}{2 m} \, \beta \, \calO_2 \,,
\qquad
\calO_2 = - \frac{(\vec \alpha \cdot \vec p)^3}{3 m^2} 
+ \frac14 \, \left\{ \frac{r_s}{r}, \vec\alpha \cdot \vec p \right\} \,.
\end{equation}
The transformation is calculated as
$H_{\rm FW} = \ee^{\ii S_2} \, H_2 \, \ee^{-\ii S_2}$.
Taking notice of the well-known identity 
$\vec p^{\,2} \left( \frac{1}{r} \right)
= 4 \, \pi \,\delta^{(3)} \left( \vec r \right)$,
the Foldy--Wouthuysen transformation gives the result,
\begin{align}
\label{HFW}
& H_{\rm FW} =
 \beta \, \left( m + \frac{\vec p^{\,2}}{2 m} - 
\frac{\vec p^{\,4}}{8 m^3} \right) 
- \beta \, \frac{m \, r_s}{2 \, r} 
\\[1.0ex]
& \; + \beta \, \left( 
- \frac{3 r_s}{8 m} \, 
\left\{ \vec p^{\,2}, \frac{1}{r} \right\}
+ \frac{3 \pi r_s}{4 m} \delta^{(3)}(\vec r) \,
+ \frac{3 r_s}{8 m} \, \frac{\vec\Sigma \cdot \vec L}{r^3} \right) \,.
\nonumber
\end{align}
The very last term on the right-hand side of Eq.~\eqref{HFW} 
describes the gravitational spin-orbit coupling;
it is in agreement with the classical result for the 
interaction of a spinning classical particle 
with the gravitational field, which is otherwise known as
the geodesic precession or Fokker precession~\cite{Fo1920,Kh2001,Kh2008}.
When comparing the spin-orbit term with Eq.~(26) of Ref.~\cite{Kh2008},
which has a prefactor of $3/2$ instead of $3/8$,
one should take note that the spin operator carries 
a factor one half ($\vec S = \frac12 \, \vec\Sigma$),
and there is an additional factor 2 in the definition of the Schwarzschild 
radius. The prefactor $\beta$ in Eq.~\eqref{HFW} describes the particle-antiparticle
symmetry, which cannot be obtained based on classical 
considerations~\cite{Fo1920,Kh2001,Kh2008}.
The result~\eqref{HFW}
thus offers a straightforward physical interpretation,
as follows: First, we have the usual relativistic corrections 
to the kinetic energy. The second term on the 
right-hand side of Eq.~\eqref{HFW} is  the gravitational potential,
duly decorated with a $\beta$ prefactor which ensures that
antiparticles are attracted.
The third term consists of a kinetic correction to the gravitational 
coupling, and a Darwin (zitterbewegung) term which is not experimentally relevant
because it is located at the center of the planet, i.e.,
inside the Schwarzschild radius. However, for completeness,
we should include this term. Finally, the spin-orbit coupling term 
commutes with the Dirac angular operator 
$K = \beta ( \vec \Sigma \cdot \vec L + 1)$ 
and with the total angular momentum (orbital$+$spin).

Let us conclude the discussion by pointing out a subtlety.
One might think  that, if the total angular momentum 
operator commutes with the Hamiltonian,
then it should automatically commute with the 
Foldy--Wouthuysen transformed Hamiltonian.
However, that is not the case.
If $A$ is an operator that commutes with the 
Hamiltonian, $[H_{\rm DS}, A] = 0$, and $H_{\rm FW} = U \, H_{\rm DS} \, U^{-1}$, then 
then the transformed operator $A_{\rm FW} = U \, A \, U^{-1}$ 
commutes with $H_{\rm FW}$. 
This can be seen as follows,
\begin{equation}
\label{unitary}
[ H_{\rm FW}, A_{\rm FW}] = 
U \, [ H_{\rm DS}, A] \, U^{-1} = 0 \,,
\end{equation}
So, if $\vec J$ commutes with $H_{\rm DS}$, then this does not 
automatically mean that $\vec J$ commutes with $H_{\rm FW}$.
Let us recall that the total angular momentum $\vec J$ and 
the Dirac angular operator $K$ are defined as follows,
\begin{align}
\vec J =& \; \vec L + \tfrac12 \, \vec\Sigma \,,
\qquad
K = \beta \, \left( \vec \Sigma \cdot \vec L + 1 \right) \,.
\end{align}
We can establish the following commutator relations,
\begin{subequations}
\label{all_commutes}
\begin{align}
[H_{\rm DS}, K ] =& \; 0 \,,
\qquad
[H_{\rm DS}, \vec J ] = \vec 0 \,.
\\
[H_{\rm FW}, K ] =& \; 0 \,,
\qquad
[H_{\rm FW}, \vec J ] = \vec 0 \,.
\end{align}
\end{subequations}
In view of~\eqref{unitary}, it is
not a triviality to separately establish that both 
$\vec J$ and $K$ commute with $H_{\rm DS}$ and $H_{\rm FW}$, individually,
but the relations hold.
Eigenfunctions of $H_{\rm DS}$ and of $H_{\rm FW}$ 
are eigenstates of $\vec J$ and $K$, i.e., the spinor 
components of the eigenfunction are $\chi_{\varkappa \mu}(\hat r)$ 
(upper component) and 
$\chi_{-\varkappa \mu}(\hat r)$ (lower component),
and they have the same $j = |\varkappa|-1/2$.

%
%
\subsection{Transformation of the Transition Current}

The main result derived in the current work 
concerns the Foldy--Wouthuysen Hamiltonian $H_{\rm FW}$
derived in Eq.~\eqref{HFW}.
However, it is also interesting to investigate the gravitational
corrections to the electromagnetic transition current.
The transition current operator for the emission of 
transverse photons in flat space is given by the
matrix-valued expression $j^i = \alpha^i \, \exp(\ii \vec k \cdot \vec r)$;
an illustrative discussion can be found in Refs.~\cite{BjDr1964,JePa1996,Je1996}.
The Hermitian adjoint of this operator 
is obtained by the replacement $\exp(\ii \vec k \cdot \vec r) \to
\exp(-\ii \vec k \cdot \vec r)$, i.e., by the replacement 
of a photon emission by a photon absorption process.

The Dirac--Schwarzschild Hamiltonian~\eqref{HDS} is coupled
to an external electromagnetic field by the replacement
$\vec p \to \vec p - e\, \vec A$, where $\vec A$ is the 
vector potential.
The interaction Hamiltonian is $H_{\rm int} = -\vec j \cdot \vec A$.
So, with relativistic gravitational coupling included, 
the transition current reads as 
\begin{equation}
\label{ji}
j^i = \frac12 \, \left\{ 1 - \frac{r_s}{r}, 
\alpha^i \, \exp(\ii \vec k \cdot r) \right\} 
\end{equation}
We now employ the multipole expansion 
\begin{equation}
\alpha^i \, \exp(\ii \vec k \cdot r) \approx
\alpha^i + \alpha^i \, (\ii \vec k \cdot \vec r) -
\frac12 \alpha^i (\vec k \cdot r)^2
\end{equation}
in Eq.~\eqref{ji}. A subsequent calculation of the 
Foldy--Wouthuysen transformation $j_{\rm FW}^i = U \, j^i \, U^{-1}$ 
of the transition current
[with $U = \exp(\ii S_2) \, \exp(\ii S_1)$] gives the result,
\begin{align}
\label{jFW}
j_{\rm FW}^i = & \;
\frac{p^i}{m} - \frac{p^i \, \vec p^{\,2}}{2 m}
- \frac{\ii}{2 m} \left( \vec k \times \vec\sigma \right)^i
+ \frac12 \, \left\{ \frac{p^i}{m}, \, (\ii \vec k \cdot \vec r) \right\}
\nonumber\\[1.0ex]
& \; 
-\frac{1}{4} \, \left\{ (\vec k \cdot \vec r)^2, \frac{p^i}{m} \right\}
+ \frac{1}{2 m} \left( \vec k \cdot \vec r \right) \, (\vec k \times \vec\sigma)^i 
\nonumber\\[1.0ex]
& \; 
- \frac{3}{4} \, \left\{ \frac{p^i}{m}, \frac{r_s}{r} \right\} 
+ \frac{r_s}{2 r} \frac{(\vec\sigma \times \vec r)^i}{m \, r^2} 
\nonumber\\[1.0ex]
& \; 
- \frac12 \, \left\{ \left( \ii \vec k \cdot \vec r \right), 
\left\{ \frac{p^i}{m}, \frac{r_s}{r} \right\} \right\} 
+ \frac{3 \ii r_s}{4 r} \frac{( \vec k \times \vec \sigma )^i}{m} 
\nonumber\\[1.0ex]
& \; + \frac{1}{4} \,
\left\{ \frac{r_s}{r} \, (\ii \vec k \cdot \vec r), 
\frac{p^i}{m} \right\} \,.
\end{align}
In addition to the canonical corrections to the 
relativistic transition current~\cite{BjDr1964,Je1996,JePa1996}
(kinetic corrections and magnetic coupling),
this result contains a gravitational kinetic correction,
and gravitational corrections to the 
magnetic coupling.

%
%
\section{Conclusions}
\label{conclu}

In this paper, we investigate the nonrelativistic limit of the gravitationally
coupled Dirac Hamiltonian for a Dirac particle bound to the gravitational field
of a planet. In order to calculate the relativistic corrections in the
gravitational field, we carry out a Foldy--Wouthuysen transformation with
relativistic corrections up to fourth order in the 
momenta and up to fourth ``combined'' order in the momenta 
and the gravitational interaction, i.e., we include 
terms of third order in the momenta
($\vec p^{\,4}$) and first order in the gravitational constant $G$. Within our
expansion, we have $|\vec p| \sim \xi \, m$, where  $m$ is the fermion mass.
Our calculations are thus valid up to the orders $\xi^4$ and
$\chi \, \xi^3$, where $\chi = r_s/r$ is the gravitational expansion parameter.
We verify that the equivalence principle holds for the gravitational
interaction of particles and antiparticles, based on an inspection of the
functional form of the relativistic corrections, which are all proportional to
the $\beta \equiv \gamma^0$ matrix [see Eqs.~\eqref{dirac_rep}
and~\eqref{HFW}].  A coupling term
proportional to $\vec \Sigma \cdot \vec g$, where $\vec g$ 
is the acceleration due to gravity, 
would otherwise break parity and the equivalence principle.
The conceivable existence of such a term had been discussed
at various places in the literature; we find that it vanishes.

One may wonder if the spin operator in our gravitational spin-orbit 
coupling really describes the physical spin of the electron.
In order to shed light on this question, we first observe that 
the total angular momentum 
$\vec J  = \vec L + \vec S = \vec L + \frac12 \, \vec \Sigma$ of the electron
commutes with the free Dirac Hamiltonian
(a particularly clear exposition can be 
found in Sec.~11.3 of Ref.~\cite{De2005}), as well as the 
Dirac--Coulomb Hamiltonian~\cite{SwDr1991a,SwDr1991b,SwDr1991c} 
and the Dirac--Schwarzschild Hamiltonian [see Eq.~\eqref{all_commutes}].
One may attempt to construct a ``relativistic''
spin operator $\vec Z$ [see Eq.~(22) of Ref.~\cite{Ry1999}] 
whose square $\vec Z^2$ ``alone'' (without any 
addition of orbital angular momentum) commutes with the fully relativistic 
Dirac Hamiltonian~\cite{Ry1999,BaAhKeGr2013}. 

In nonrelativistic limit accessed by the Foldy--Wouthuysen transformation,
all ``relativistic'' spin operators discussed so far in the literature
approach the limiting form $\vec S = \frac12 \, \vec \Sigma$ 
[see Eq.~(16) of Ref.~\cite{Ry1999} and the text following Eq.~(4) of 
Ref.~\cite{BaAhKeGr2013}]. We can thus uniquely identify our 
result as the gravitational spin-orbit coupling term
corresponding to the classical geodesic precession
(Fokker precession) of a spinning 
object in the gravitational field~\cite{Fo1920},
formulated within quantum mechanics and respecting the 
particle-antiparticle symmetry (prefactor $\beta$).
The spin of the electron precesses due to the coupling
to the gravitational field, without violating parity symmetry.
We also find the gravitational analogue
of the zitterbewegung term, in the Foldy--Wouthuysen transformed 
Dirac--Schwarschild Hamiltonian~\eqref{HFW}.
The result is complemented by the 
gravitational corrections to the transition current~\eqref{jFW}.

%
%
\section*{Acknowledgments}

The authors acknowledge helpful conversations with T. N. Moentmann,
Professor~M.~I.~Eides as well as support by the National Science Foundation
(Grant PHY--1068547) and by the National Institute of Standards and Technology
(precision measurement grant).

\appendix

%
%
\section{Foldy--Wouthuysen of the Free Dirac Hamiltonian}
\label{appa}

It is instructive to carry out the Foldy--Wouthuysen 
transformation of the free Dirac Hamiltonian, in 
exactly the same two-step approach, with two iterative
transformations, as described in Ref.~\cite{BjDr1964}
for the Dirac--Coulomb Hamiltonian (keeping in mind that 
the Foldy--Wouthuysen transformation of the free Dirac Hamiltonian 
can otherwise be carried out in a single step).
The free Dirac Hamiltonian is given as,
\begin{equation}
H_{\rm FD} = H_1 = \vec \alpha \cdot \vec p + \beta m \,.
\end{equation}
For the first Foldy--Wouthuysen transformation, we have
\begin{equation}
\label{SS1}
S_1 = -\frac{\ii}{2 m} \, \beta \, \calO_1 \,,
\qquad
\calO_1 = \vec\alpha \cdot \vec p \,.
\end{equation}
The transformation is calculated as
\begin{equation}
H_2 = \ee^{\ii S_1} \, H_1 \, \ee^{-\ii S_1} \,.
\end{equation}
This leads to the result
\begin{equation}
H_2 = \beta \left( m + \frac{\vec p^{\,2}}{2 m} - 
\frac{\vec p^{\,4}}{8 m^3} \right) - \frac{(\vec \alpha \cdot \vec p)^3}{3 m^2} \,.
\end{equation}
For the second Foldy--Wouthuysen transformation, we have
\begin{equation}
S_2 = -\frac{\ii}{2 m} \, \beta \, \calO_2 \,,
\qquad
\calO_2 = - \frac{(\vec \alpha \cdot \vec p)^3}{3 m^2} \,.
\end{equation}
The transformation is calculated as
\begin{equation}
H_{\rm FW} = \ee^{\ii S_2} \, H_2 \, \ee^{-\ii S_2}\,. 
\end{equation}
The transformed Hamiltonian is 
\begin{equation}
H_{\rm FW} = \beta \, \left( m + \frac{\vec p^{\,2}}{2 m} - 
\frac{\vec p^{\,4}}{8 m^3} \right)\,. 
\end{equation}
This result corresponds to our Eq.~\eqref{HFW}
is the limit of vanishing $r_s$, as it should.

All unitary transformations discussed here respect 
the basic symmetries of the Dirac Hamiltonian such as 
parity. This is essential; as an extreme example, let us 
briefly supplement the discussion by considering the 
nonrelativistic free Schr\"{o}dinger Hamiltonian 
$H_0 = \vec p^{\,2}/(2 m)$ and the 
unitary transformation $U = \exp(\ii \vec A \cdot \vec r)$,
where $\vec A$ is a constant vector. Then,
$H'_0 = U \, H_0 \, U^\plus = (\vec p - \vec A)^2/(2 m)$.
Upon binomial expansion, one obtains a term proportional to 
$\vec A \cdot \vec p$, which breaks parity.
However, the parity-breaking term in $H'_0$ is an artefact generated 
by the parity-breaking unitary transformation.
As a further illustrative remark,
let us consider the the transformation $S_1$,
given in Eq.~\eqref{SS1}, under parity,
\begin{align}
S_1 =& \; -\frac{\ii}{2 m} \, \beta \, \vec \alpha \cdot \vec p  
\nonumber\\
\mathop{\to}^{\mathcal P} & \;
\beta \, \left[ -\frac{\ii}{2 m} \, \beta \, \vec \alpha \cdot (-\vec p) \right] \,
\beta =
\frac{\ii}{2 m} \, \vec \alpha \cdot \vec p \; \beta = S_1 \,.
\end{align}
By construction, the iterative Foldy--Wouthuysen transformations
discussed here respect parity symmetry,
and so does the final result given in Eq.~\eqref{HFW}.


\begin{thebibliography}{10}

\bibitem{CoOvWe1975}
R. Colella, A.~W. Overhauser, and S.~A. Werner, Phys. Rev. Lett. {\bf 34},
  1472  (1975).

\bibitem{BoWr1983}
U. Bonse and T. Wroblewski, Phys. Rev. Lett. {\bf 51},  1401  (1983).

\bibitem{BoWr1984}
U. Bonse and T. Wroblewski, Phys. Rev. D {\bf 30},  1214  (1984).

\bibitem{Iv1969a}
O.~S. Ivanitskaya, {\em \relax{Extended Lorentz transformations and their
  applications (in Russian)}} (Nauka i Technika, Minsk, USSR, 1969).

\bibitem{Iv1969b}
O.~S. Ivanitskaya, {\em \relax{Lorentzian basis and gravitational effects in
  Einstein’s theory of gravity (in Russian)}} (Nauka i Technika, Minsk, USSR,
  1969).

\bibitem{Mi1969}
N.~V. Mitskevich, {\em \relax{Physical fields in general relativity (in
  Russian)}} (Nauka, Moscow, 1969).

\bibitem{Go1985}
A.~K. Gorbatsievich, {\em \relax{Quantum mechanics in general relativity. Basic
  principles and elementary applications}} (Nauka i Technika, Minsk, 1985).

\bibitem{IvMiVl1985}
O.~S. Ivanitskaya, N.~V. Mitskievic, and Y.~S. Vladimirov,  in {\em
  \relax{Proceedings of the 114th Symposium of the International Astronomical
  Union held in Leningrad, USSR, May 1985}}, edited by J. Kovelevsky and V.~A.
  Brumberg (Kluwer, Dordrecht, 1985), pp.\ 177--186.

\bibitem{PeRi1987vol1}
R. Penrose and W. Rindler, {\em \relax{Spinors and space-time --- Vol. 1}}
  (Cambridge University Press, Cambridge, UK, 1987).

\bibitem{PeRi1987vol2}
R. Penrose and W. Rindler, {\em \relax{Spinors and space-time --- Vol. 2}}
  (Cambridge University Press, Cambridge, UK, 1987).

\bibitem{BrWh1957}
D.~R. Brill and J.~A. Wheeler, Rev. Mod. Phys. {\bf 29},  465  (1957).

\bibitem{Bo1975prd}
D.~G. Boulware, Phys. Rev. D {\bf 12},  350  (1975).

\bibitem{SoMuGr1977}
M. Soffel, B. M\"{u}ller, and W. Greiner, J. Phys. A {\bf 10},  551  (1977).

\bibitem{Re2011}
V.~M. Redkov, Nonlinear Phenomena in Complex Systems {\bf 7},  250  (2011).

\bibitem{Ye2011}
J. Yepez, {\em Einstein's vierbein field theory of curved space}, e-print
  arXiv:1106.2037v1 [gr-qc].

\bibitem{Je2013}
U.~D. Jentschura, Phys. Rev. A {\bf 87},  032101  (2013), [Erratum Phys.~Rev.~A
  {\bf 87}, 069903 (2013)].

\bibitem{HeNi1990}
F.~W. Hehl and W.-T. Ni, Phys. Rev. D {\bf 42},  2045  (1990).

\bibitem{Pe1978}
A. Peres, Phys. Rev. D {\bf 18},  2739  (1978).

\bibitem{Ob2001}
Y.~N. Obukhov, Phys. Rev. Lett. {\bf 86},  192  (2001).

\bibitem{FoWu1950}
L.~L. Foldy and S.~A. Wouthuysen, Phys. Rev. {\bf 78},  29  (1950).

\bibitem{Ni2002}
N. Nicolaevici, Phys. Rev. Lett. {\bf 89},  068902  (2002).

\bibitem{Ob2002}
Y.~N. Obukhov, Phys. Rev. Lett. {\bf 89},  068903  (2002).

\bibitem{SiTe2005}
A.~J. Silenko and O.~V. Teryaev, Phys. Rev. D {\bf 71},  064016  (2005).

\bibitem{Si2008pra}
A.~J. Silenko, Phys. Rev. A {\bf 77},  012116  (2008).

\bibitem{ObSiTe2011}
Y.~N. Obukhov, A.~J. Silenko, and O.~V. Teryaev, Phys. Rev. D {\bf 84},  024025
   (2011).

\bibitem{GoObSh2009}
B. Goncalves, Y.~N. Obukhov, and I.~L. Shapiro, Phys. Rev. D {\bf 80},  125034
  (2009).

\bibitem{SwDr1991a}
R.~A. Swainson and G.~W.~F. Drake, J. Phys. A {\bf 24},  79  (1991).

\bibitem{SwDr1991b}
R.~A. Swainson and G.~W.~F. Drake, J. Phys. A {\bf 24},  95  (1991).

\bibitem{SwDr1991c}
R.~A. Swainson and G.~W.~F. Drake, J. Phys. A {\bf 24},  1801  (1991).

\bibitem{BjDr1964}
J.~D. Bjorken and S.~D. Drell, {\em \relax{Relativistic Quantum Mechanics}}
  (McGraw-Hill, New York, 1964).

\bibitem{BjDr1965}
J.~D. Bjorken and S.~D. Drell, {\em \relax{Relativistic Quantum Fields}}
  (McGraw-Hill, New York, 1965).

\bibitem{ItZu1980}
C. Itzykson and J.~B. Zuber, {\em \relax{Quantum Field Theory}} (McGraw-Hill,
  New York, 1980).

\bibitem{PeSc1995}
M.~E. Peskin and D.~V. Schroeder, {\em \relax{An Introduction to Quantum Field
  Theory}} (Perseus, Cambridge, Massachusetts, 1995).

\bibitem{Sr2007}
M. Srednicki, {\em \relax{Quantum Field Theory}} (Cambridge University Press,
  Cambridge, 2007).

\bibitem{Di1975}
P.~A.~M. Dirac, {\em \relax{General Relativity}} (Wiley, New York, NJ, 1975).

\bibitem{Wa1984}
R.~M. Wald, {\em \relax{General Relativity}} (University of Chicago Press,
  Chicago, IL, 1984).

\bibitem{Ch1992}
S. Chandrasekhar, {\em \relax{The Mathematical Theory of Black Holes}} (Oxford
  University Press, Oxford, UK, 1992).

\bibitem{Sc2009}
B. Schutz, {\em \relax{A First Course in General Relativity}} (Cambridge
  University Press, Cambridge, UK, 2009).

\bibitem{Pa2010genrel}
T. Padmanabhan, {\em \relax{Gravitation --- Foundations and Frontiers}}
  (Cambridge University Press, Cambridge, UK, 2010).

\bibitem{Di1996}
P.~A.~M. Dirac, {\em \relax{General Theory of Relativity}} (Princeton
  University Press, Princeton, 1996).

\bibitem{Ed1924}
A.~S. Eddington, {\em \relax{The Mathematical Theory of Relativity}} (Cambridge
  University Press, Cambridge, England, 1924).

\bibitem{Wo1999}
S. Wolfram, {\em \relax{The Mathematica Book}}, 4 ed. (Cambridge University
  Press, Cambridge, UK, 1999).

\bibitem{YoAd2012}
T.~J. Yoder and G.~S. Adkins, Phys. Rev. D {\bf 86},  116005  (2012).

\bibitem{Je1996}
U.~D. Jentschura, {\em Master Thesis: The Lamb Shift in Hydrogenlike Systems,
  [in German: Theorie der Lamb--Verschiebung in wasserstoffartigen Systemen],}
  (University of Munich, 1996, unpublished (see e-print hep-ph/0305065)).

\bibitem{Fo1920}
A.~D. Fokker, Kon. Akad. Weten. Amsterdam. Proc. {\bf 23},  729  (1920).

\bibitem{Kh2001}
I.~B. Khriplovich,  in {\em \relax{Gyros, Clocks, Interferometers$\dots$:
  Testing Relativistic Gravity in Space (Lecture Notes in Physics 562)}},
  edited by C. L\"{a}mmerzahl, C.~W.~F. Everitt, and F.~W. Hehl (Springer,
  Heidelberg, 2001), pp.\ 109--128.

\bibitem{Kh2008}
I.~B. Khriplovich, Acta Phys. Polon. B Proc. Suppl. {\bf 1},  197  (2008).

\bibitem{JePa1996}
U. Jentschura and K. Pachucki, Phys. Rev. A {\bf 54},  1853  (1996).

\bibitem{De2005}
V. Devanathan, {\em \relax{Quantum Mechanics}} (Alpha Science, Oxford, UK,
  2005).

\bibitem{Ry1999}
L.~H. Ryder, Gen. Relativ. Gravit. {\bf 31},  775  (1999).

\bibitem{BaAhKeGr2013}
H. Bauke, S. Ahrens, C.~H. Keitel, and R. Grobe, {\em What is the relativistic
  spin operator?}, e-print arXiv:1303.3862v1 [quantum-ph].

\end{thebibliography}
\end{document}